# Optical Micromanipulations based on Model Predictive Control of Thermoviscous Flows

Elena Erben[1,*], Ivan Saraev[1,*], Weida Liao[2], Fan Nan[1], Eric Lauga[2], Moritz Kreysing[1,**]

1) Institute of Biological and Chemical Systems – Biological Information Processing. Karlsruhe Institute of Technology (KIT), Germany.

2) Department of Applied Mathematics and Theoretical Physics (DAMTP), University of Cambridge, UK.

\* These authors contributed equally to this work.
\*\* Correspondence: moritz.kreysing@kit.edu

## Abstract

High-precision micromanipulation techniques, including optical tweezers and hydrodynamic trapping, have garnered wide-spread interest. Recent advances in optofluidic multiplexed assembly and microrobotics demonstrate significant progress, particularly by iteratively applying laser-induced, localized flow fields to manipulate microparticles in viscous solutions. However, these approaches still face challenges such as undesired hydrodynamic coupling and instabilities when multiple particles are brought into close proximity. By leveraging an analytical model of thermoviscous flows, this work introduces a stochastic optimization approach that selects flow fields for precise particle arrangement without relying on rule-based heuristics. Through minimizing a comprehensive objective function, the method achieves sub-micrometer alignment accuracy even in a crowded setting, avoiding instabilities driven by undesired coupling or particle collisions. An autonomously emerging "action at a distance" strategy—placing the laser scan path farther from the manipulated particles over time—exploits the $1/r^2$ decay of thermoviscous flow to refine positioning. Overall, objective function-based model predictive control enhances the versatility of automated optofluidic manipulations, opening new avenues in assembly, micromanufacturing, robotics, and life sciences.

# INTRODUCTION

The precise manipulation of colloidal particles remains an important research pursuit, with potential benefits for life sciences, biomedical research, and nanoscience.[1–8] Conventional micromanipulation techniques, including optical tweezers,[9–12] optoelectronic tweezers,[13,14] and magnetic tweezers,[15,16] rely on intricate light–matter or magnetic interactions, or both. Consequently, these methods are constrained by material dependencies, a limited manipulation range, and low throughput in living systems. Highly controllable, optically induced flows (i.e., hydrodynamic manipulation), which can be created through the movements of optically trapped microparticles[17,18] or through optothermal effects,[6,19–24] could prove an effective solution to the aforementioned issues, potentially enabling the efficient manipulation of multiple particles. The aforementioned approach is typically based on optical tweezers, which, while highly successful in many use cases, have some fundamental limitations that render them less suitable for others. Specifically, since the motion of the trapped particles is induced by the shifting of the laser spots, when the laser beam moves at a high speed, the optical gradient force may prove insufficient to balance the fluidic drag force generated by the particles' motion, especially if the refractive index difference between the particle and the medium is small.[25] Consequently, the control over a particle may be lost, particularly in viscous media, leading to the disappearance of the induced flow field. The latter techniques, specifically those based on the thermoviscous effect,[20,26] have been demonstrated to be effective for a number of applications. These include force measurements in the femtonewton range,[27] optofluidic multiplexed assembly of up to 15 particles,[28] overcoming a fundamental constraint of pump-actuated Hele-Shaw cells,[28,29] and microrobotics applications,[28,30] even within quasi-isothermal microenvironments.[31] Beyond its use for microparticle manipulation, there is a growing body of work leveraging induction of thermoviscous flows inside biological cells to investigate fundamental biological properties and processes.[32–35]

While the extended 2D thermoviscous flow field has been studied experimentally[20,28,30] and theoretically,[28,36] the previous control method remained at the level of sequential per-particle displacements, facilitated by simple, explicit heuristics. Specifically, it utilized only the straight-lined flows localized along the scan path axis for directed particle alignment. Furthermore, this rule-based approach aimed to position a single particle per laser scan path and feedback step, without considering the effect of the flow field on the other particles. While this positioning approach yielded successful manipulations for large particle spacings, it caused instabilities and convergence problems when particles became too close to each other, originating from undesired crosstalk between particles and their positioning flow fields. Despite the implementation of simple context-aware policies utilizing different scan path positions in the presence of other particles nearby,[28] the scope of manipulations remained constrained to cases involving no more than two particles in proximity to one another. Instabilities impeding convergence were particularly observed during alignment scenarios with spacings of 10 µm or less, or as consequences of random encounters of particles during alignments.

Although thermoviscous flow fields have become analytically tractable,[36,37] significant uncertainty remained regarding the potential of stepwise deterministic simulations to circumvent these convergence issues and instabilities in parallel optofluidic alignment processes. Additionally, it was previously unclear whether the model-based and presumably time-expensive search for suitable flow fields would be a viable solution for manipulating particles hydrodynamically in a medium where they are free to diffuse. In this study, we present an approach that employs model predictive control to minimize a global objective function, to enhance microparticle manipulation via predictable, optically induced flow fields. This is demonstrated by means of deterministic simulations to solve the inverse problem of finding suitable flow fields for particle alignment. We complemented these simulations by proof-of-concept experiments on time scales on which random diffusion significantly accompanies the manipulations. Our approach establishes the foundation for the control of optically induced thermoviscous flow fields via objective function optimization techniques, which can be carried out in a mathematically tractable manner. We find that simple stochastic optimization is sufficient to align particles with sub-micrometer precision via emergent manipulation strategies at a distance, quantitatively explained by analytical modeling, and while overcoming previously observed limitations of close particle spacings. Our results demonstrate the first application of the analytical model of thermoviscous flows to enhanced colloid manipulation. This constitutes a significant conceptual advancement which we expect will increase the versatility of the technique and open the door to the use of more advanced control algorithms and objective functions for applications in assembly, microfabrication, the life sciences, and robotics.

## RESULTS

### Simulations: Sufficiency of Thermoviscous Flows to Align Particles via Model Predictive Control

Thermoviscous flows arise from the repeated movement of a heating infrared (IR) laser spot through a thin liquid film, enabled by a specialized microscope setup (Figure 1a). It has been shown that these flows can propel particles in a fluid at speeds of up to 150 μm s$^{-1}$,[20] decaying spatially as $1/r^2$ in the far field.[30] Here, we demonstrate that these flows offer a way to align particles into 2D assemblies by optimizing a global objective function (Figure 1b).

To this end, we leverage a full analytical model of the 2D flow field[36] (Figure S1a, Supporting Information) to solve the inverse problem of identifying flow fields that align particles with their intended destinations. Specifically, a random search approach (Figure 1c) was employed to randomly generate a candidate flow field by varying only the position and angular orientation of the laser scan path while keeping other parameters (i.e., scan path length, scan frequency, and laser power) constant. Subsequently, the flow field is evaluated using forward simulation to determine whether it improves the objective function beyond a percentage threshold of typically a few percent. Once a satisfactory flow field has been identified, it is applied, and the search begins anew based on the updated particle positions. We are thereby approaching the alignment problem as an objective function optimization problem that considers all the particles' positions collectively, rather than on a particle-by-particle basis, which previously showed described limitations.[28] Moreover, by taking advantage of the full theoretical flow field throughout space,[36] our algorithm stands in sharp contrast to previous work reliant on the simplified rule that particles move in the opposite direction to the laser scan locally.[20,28]

### Simulations Penalizing the Root-Mean-Square Distance Converge

We find that simulations converge by minimizing an objective function given by the root-mean-square (RMS) spacing of particles from their destinations. Here, we present the most challenging scenario that was evaluated: nine particles approaching a dense 3 × 3 grid (Figure 2a; Video S1, Supporting Information). To achieve alignment, the objective function is required to reduce beyond a percental threshold factor (typically between 90% and 98%). Consistent with this notion of an exponential decay process, we observe effective convergence already after tens of steps (Figure 2c).

Strikingly, even in this challenging scenario involving nine particles in close proximity, we did not observe issues associated with undesired hydrodynamic coupling, such as oscillatory behavior or stalling convergence, which we frequently encountered when attempting to align more than two particles using the previous rule-based iterative particle-by-particle alignment strategy.[28] Moreover, this approach did not require the explicit formulation of context-aware decision-making rules, which we implemented in our previous approach through the use of pushing or pulling flows.[28]

Beyond the approach of the particles to close set points, we also observed a similar good convergence for the expansion of the lattice, i.e., the transition to an increased spacing (Figure 2b; Video S1, Supporting Information).

**Objective Functions Including Interparticle Interaction Terms**

In the employed framework, particles are assumed to move with the flow as point-like entities, with minimal impact on the generated flow field itself. However, a limitation to this viewpoint is that particles might be hindered in their motion by other particles, as the simulation does not account for their finite size and excluded volume interactions.

To address this issue, we expanded the objective function to consider not only the pairwise RMS displacement of particles from their destinations, but also the distances between all particle pairs. An effective Lennard-Jones (LJ) type potential with strong repulsion (scaling with the interparticle distance as $1/|\boldsymbol{p}_i - \boldsymbol{p}_j|^{12}$) was introduced, taking effect below a critical interparticle distance $2d$. This results in the introduction of 36 additional terms to the objective function in the case of nine particles. These mixed terms effectively prevent collisions by maintaining a minimal distance between particles throughout the alignment (Figure 2d), not only when terminally positioned. This allows particles to navigate around each other during alignment. We are thereby conceptually transitioning from the optimization of individual behavior in a multiparticle system to emergent collective dynamics.

The results of these simulations indicate that particles can be aligned in a mathematically tractable manner using an objective function approach, without the need for rule-based policies. These simulations further demonstrate the sufficiency of thermoviscous flows for particle alignment even in challenging scenarios. Although other physical effects, such as thermophoresis,[19] thermo-osmosis,[6] or temporary optical trapping of beads, may occur during the generation of thermoviscous flows, our simulations indicate that these alternative effects are not necessary for successful alignment. The sufficiency of thermoviscous flows is of particular interest, given that this optically generated flow fields are known to possess particularly strong localization (with a $1/r^2$ decay in the far field), rendering them highly relevant for precision particle manipulation.

**Successful Alignment in Experiments**

Having demonstrated the sufficiency of thermoviscous flows for particle alignment through deterministic simulations, we performed experiments to validate our approach under realistic conditions, where factors such as diffusion, brief interruptions to execute calculations, and deviations from the analytical flow field model can exert a considerable influence. To this end, we used the aforementioned simulations to predict a successively improving step. To account for random displacements, for example, those resulting from diffusion, we employed a feedback loop (Figure 1c), collecting the new particle positions from the camera image after applying a predicted flow field.

Using 3 µm polystyrene (PS) beads in $H_2O$–glycerol mixtures in 5 µm thin chambers, we found that particle alignment via thermoviscous flows using objective function optimization can successfully align colloids in experiments (Figure 3a; Video S2, Supporting Information). Initially, we observed deviations in absolute speed between the analytical model and the PS particles. We attribute these deviations to the spatial extent of the particles which subjects them not only to the mid-plane velocity modeled for point-like particles but also to lower speeds closer to the no-slip chamber walls.

Consequently, we had to scale up the modeled velocities by an empirical factor (see the "Methods"). Any remaining discrepancies between model and experiment, such as sudden particle displacements due to optical interactions upon direct laser exposure, were corrected by the feedback loop. Such direct laser exposure of the manipulated particles can be prevented by imposing the additional constraint of a minimum distance between the laser scan path and the particles (see Figure S4 in the Supporting Information). In addition to unwanted optical interactions, this could also be used to prevent exposure of sensitive probe particles to the peak temperature induced by the laser beam.

This method does not require prior specification of the trajectories or context-dependent policies for generating a flow field in crowded settings. One noteworthy attribute of the leveraged random search approach is the emergence of unanticipated particle trajectories and "alignment strategies" (Figure 3b). The objective function (RMS distance) decreases near-monotonically with time (Figure 3c). However, the distance of the individual particle to its target can increase temporarily. The decrease of the RMS distance is stalled during the waiting periods when no flow fields were applied (see the "Methods"). Trapping of the particles at the target location is achieved with a moderate precision of roughly 350 nm (Figure 3d). Decreasing the feedback sampling period should enable a higher trapping precision.

**Precision Action at a Distance**

Additionally, these alignment experiments revealed that, as the alignment progressed, the selected scan path would be placed further and further away from the particles themselves (Figure 3e). This emergent behavior was observed to be consistent across multiple repetitions of the experiment (Figure S3, Supporting Information). In the final stages of the alignment process, such actuation at a distance enables finer control of the particles, due to lower thermoviscous flow velocities induced at their location (for fixed laser power). This may be explained quantitatively via a scaling argument. The speed of particles decays spatially as $1/r^2$ far from the scan path. Therefore, requiring an improvement by a specified percentage of the RMS distance $L$ of particles to their targets (i.e., the objective function) yields a power law for the characteristic distance $r$ of the scan path from the nearest particle as $r \propto L^{-1/2}$. This finding underscores that particles do not require exposure to the laser beam or heat stimulus; in fact, avoiding it is advantageous, as it results in greater precision. When we incorporate prefactors into our theoretical argument (see the "Methods"), we may also interpret the resulting characteristic distance of the scan path from the nearest particle as an approximate upper bound. Furthermore, this emergent behavior contrasts with the preceding rule-based methodology, where the distance between the scan path and the particles was kept constant throughout the alignment process, effectively using only the near-field thermoviscous flow to manipulate the particles. The approach presented here instead makes use of the far field as well, to enable precise and parallel manipulation. This upper bound for the distance of the scan path from the particles could be utilized in future development of this method to increase computational efficiency by limiting the space from which scan paths are sampled.

**Model Predictive Control Achieves Complex Particle Arrangements without Hydrodynamic Instabilities**

We next investigated how our objective function-based approach performs for crowded target arrangements that present a significant challenge to the previous rule-based algorithms,[28] i.e., with many particles and closely spaced targets. For this purpose, we experimentally examined the 3 × 3 grid arrangement for which our earlier simulations converged (Figure 4a; Video S3, Supporting Information). In the experiment, we found not only robust and reliable convergence of the objective function (RMS distance to target, Figure 4b), but also a terminal positioning precision of 1.2–1.5 µm in steady state (Figure 4c). Conversely, the alignment of such target arrangements using the previous rule-based approach[28] was frequently characterized by pronounced oscillation or stalemate-like repetitive behavior (see Video S4 in the Supporting Information), which delayed and, in extreme cases, prevented convergence. We attribute this to the fixed position and orientation of the scan path (relative to a particle and its target) calculated by that algorithm, based only on knowledge of the transport direction collinear to the scan path. This led to increased hydrodynamic coupling in arrangements where particles were densely packed. In contrast, here, during multiple repetitions of our model predictive control approach, we did not observe such problems.

Furthermore, this example demonstrates that the number of controlled particles can be readily increased in this microfluidic manipulation technique using thermoviscous flows, even beyond the limitations of conventional microfluidic flow chambers (Hele-Shaw cells). In such chambers, it has been shown that the number of mobile particles that can be manipulated simultaneously is theoretically limited to six.[29] We have shown that automated particle positioning via thermoviscous flows overcomes this limitation, both via an iterative, particle-by-particle approach in our previous work[28] and here via model predictive control, which further prevents instabilities when handling closely spaced particles that were encountered with the former method.

**DISCUSSION AND OUTLOOK**

Thermoviscous flows offer numerous advantages for the manipulation of colloids. Optical generation of these flows is possible at any position within a sample by the repeated movement of an IR laser spot along a linear trajectory. In contrast to other manipulation methods, microfluidic techniques are not constrained by material properties or the potential for harm to samples. As thermoviscous flows are independent of the viscosity of the medium, they appear to be particularly well suited for applications where force-driven or phoretic approaches are too slow, or where a viscosity-mediated reduction of diffusion is of interest. Furthermore, the far-field spatial decay of the induced net thermoviscous flow as $1/r^2$ offers flexibility and precision in particle manipulation, without direct laser exposure and without depending on particle material properties.

In this work, we demonstrate that the analytical model of thermoviscous flows[36] can be employed to control particles using a model predictive control method. Equipped with the full, spatially varying flow field, this model-based approach takes advantage of both the near and far fields of the flow and is hence able to discover scan paths that position multiple colloids simultaneously. This is in stark contrast to the previous rule-based approach for automated thermoviscous flow control, which exploited only the essentially 1D transport along the line of the scan path, while the motion induced far from the scan path was a side effect that was detrimental to alignment goals by disrupting the positioning of other particles. Consequently, the model-based approach naturally selects flow fields that reliably circumvent instabilities caused by hydrodynamic coupling, which were observed in the previous rule-based approach.[28] Through simulations, we further found that the alignment can be attributed exclusively to thermoviscous flows, implying that other thermally induced effects that might accompany flows,[6] particularly in low-viscosity fluids, are not necessary for precise particle manipulation.

The model predictive control approach presented here constitutes a substantial advance in the transition from rule-based to objective function-based descriptions. The current approach allows for the formulation of one or more concrete objective(s) that can be weighed against each other, such as the repulsive term employed here to incorporate the real-world physical constraint of excluded volumes. Complementing this, the capability to achieve these goals arises from incorporation of the analytical description of flow throughout space into flow-field selection. While the current approach is solely oriented toward optimizing the objective function, it is important to note that future research could also explore the potential to optimize the particles' path to their destinations. In addition to computationally encoding the goal of alignment, objective functions could be used to specify supplementary objectives or constraints for the process. This could include the prevention of interactions between reactive particles by maintaining a minimum distance between them or imposing constraints on generated temperature fields to prevent damage to heat-sensitive samples, such as cells, or to prevent unwanted optical trapping effects. Furthermore, the strategic engineering of objective functions could also result in a reduction of the convergence time. In experimental scenarios where the relative position of particles is of greater consequence than their absolute position, the utilization of descriptive objective functions instead of functions based on particle–target distances could reduce the amount of particle repositioning required.

Furthermore, our experimental and theoretical findings indicate that precision positioning during the final phase of the alignment process is best achieved through flow field actuation at a greater distance, due to the smaller particle displacements induced. Quantitatively, in this regime, the distance of a selected scan path to the nearest particle is approximately proportional to the inverse square root of the RMS distance of particles to their targets. This emerging "strategy" emphasizes the potential of manipulating particles remotely, thereby avoiding direct laser exposure. This demonstrates how behaviors that are not explicitly programmed into the approach can emerge in experiments and assays that seek their path to fulfill a desired objective in a stochastic, partially unpredictable manner.

While demonstrating effectiveness for alignment in all the tested scenarios, there is considerable potential to improve the speed and the precision of the alignment process in the future. The current feedback sampling period of 1.67 s is notably longer than that of other methods[38] and the typical timescales of diffusion in the media we use. A reduction in the sampling period should result in enhanced positioning precision and a more rapid alignment convergence by limiting the time for which a particle can freely diffuse. Theoretically, a 100-fold faster flow field search would reduce the diffusional displacement by a factor of 10, thereby rendering precision in the range of tenths of nanometers realistic. Within the control framework presented here, a lower sampling period could be achieved through the optimization of hardware components, such as increasing the number of central processing unit (CPU) cores and graphics processing units (GPUs), or use of field-programmable gate arrays (FPGAs). Additionally, the software architecture can be enhanced with implementation of lookup tables for particle trajectory modeling, asynchronous processing and adaptive timing control, and constraints on the space from which to sample scan paths based on the theoretical limit. Transitioning from a computationally costly random search approach toward more elaborate gradient descent methods will facilitate accelerated convergence and enhanced precision via path optimization in systems where diffusion is the limiting factor. It is also imperative to acknowledge the potential of path optimization to facilitate much more difficult alignments where the current algorithm may not converge.

Beyond the immediate scope of particle manipulation, we see potential applications of our findings in various fields. In the context of cell manipulations, our technique could be employed to induce cell contacts, or to position cells without the need for genetic intervention.[39,40] In the realm of cell–pathogen interactions and viral biology, these methods could provide precise control over experimental conditions.[41] Furthermore, these techniques could facilitate the induction of events that otherwise occur only randomly and infrequently. This includes the alignment of cells with on-chip devices and printed structures, as well as the effective interfacing of conventional microfluidic systems, as demonstrated.[42–44]

## METHODS

**Experimental Setup for Generation of Thermoviscous Flows**

The experimental setup comprised a custom-built infrared (IR) laser scanning unit and an inverted microscope, designed for both laser projection and fluorescence imaging, as outlined by Mittasch et al.[32] The infrared unit utilized a fiber-based IR Raman laser (1455 nm wavelength, continuous-wave mode, 20 W maximum power, CRfl-20-1455-OM1, Keopsys). A 2D acousto-optic deflector (AOD, AA.DTSXY-A6-145, Pegasus Optik) enabled fast and accurate scanning of the IR laser beam. The microscope utilized (IX83, Olympus) was equipped with both bright-field and fluorescence imaging optics, as well as a spinning-disk confocal unit (CSU-X1, Yokogawa). This facilitated imaging of fluorescent and nonfluorescent samples with reduced background interference and precise z-plane scanning. IR laser position and camera imaging were controlled via a LabVIEW program on an Intel Xeon E5-2630 workstation (6 cores at 2.6 GHz, 16 GB RAM). The actuation of the AODs was achieved via a PCI express card (PCIe-6363, National Instruments) and additional electronics for signal transformation, as detailed by Mittasch et al.[32]

**Random Search Algorithm for Stochastic Optimization**

In this work, a random search algorithm was used to identify a sequence of flow fields that optimize a global objective function characterizing the alignment of particles to targets (see Figure 1c). In each iteration, the current particle position was initially detected. In simulations, the initial particle positions for each time step were calculated using the same analytical model that was used to select the flow field to be applied. In experiments, the current particle positions at the beginning of each time step were derived from a live camera image of the sample. Particles were then assigned to their intended target positions. To this end, the Hungarian algorithm, also known as the Kuhn–Munkres algorithm,[45] was used to find the particle–target pairs that minimize the sum of the squared distances between them. For this purpose, the package "munkres" for Python, created by Brian Clapper, was used.[46] Based on this assignment, the objective function was calculated. Subsequently, a random scan path was generated by sampling a random center coordinate (uniformly distributed across a predefined window) and a random angle for it (uniformly distributed across $2\pi$). The movement of particles due to this laser scan was predicted using the analytical model of the 2D flow field.[36] If the value of the objective function is reduced by a predefined percentage (the "improvement" value), the scan path is deemed acceptable and applied. Otherwise, a new scan path was sampled randomly.

**Analytical Model for Particle Displacement Due to Laser Scanning**

In order to model the 2D flow field generated by each laser scan, the analytical model developed by Liao et al.[36] was employed. According to this model, the scanning can be thought of as a discrete process, whereby each laser pass induces fluid flow and, consequently, net displacement of the particles. The displacement of a particle resultant from a single scan depends on its position relative to the coordinates of the laser scan, the characteristic radius of the heat spot $a$, the peak temperature change induced by the laser $\Delta T_0$, the thermal expansion coefficient $\alpha$, and the thermal viscosity coefficient $\beta$. The displacement $\Delta \boldsymbol{X}$ of a particle at position $(X_0, Y_0)$ due to a single laser scan along a path of length $2l$, which is centered around the coordinate grid center and scanned along the $x$-axis, can be calculated for a Gaussian temperature profile according to Liao et al.[28,36] as follows

$$\Delta \boldsymbol{X}(X_0, Y_0) = \frac{3}{2} \alpha \beta \Delta T_0^2 \int_{-t_0}^{t_0} \boldsymbol{u}_{1,1}(X_0, Y_0, t) \, dt \, , \quad (1)$$

with

$$\boldsymbol{u}_{1,1}(x, y, t) = A(t)^2 U \left\{ \boldsymbol{e}_x \left( \frac{a^2}{4r^2} - \frac{a^2(x - Ut)^2}{2r^4} + \left[ -\frac{a^2}{2r^2} + \frac{a^2(x - Ut)^2}{r^4} \right] \exp(-r^2/2a^2) \right. \right.$$
$$+ \left[ \frac{a^2}{4r^2} - \frac{a^2(x - Ut)^2}{2r^4} \right] \exp(-r^2/a^2) - \frac{1}{4} E_1(r^2/2a^2) + \frac{1}{4} E_1(r^2/a^2) \right)$$
$$\left. + \boldsymbol{e}_y \left[ -\frac{a^2(x - Ut)y}{2r^4} + \frac{a^2(x - Ut)y}{r^4} \exp(-r^2/2a^2) - \frac{a^2(x - Ut)y}{2r^4} \exp(-r^2/a^2) \right] \right\} , \quad (2)$$

$$E_1(z) \equiv \int_z^\infty \frac{\exp(-s)}{s} \, ds \, , \quad (3)$$

and

$$A(t) = \cos^2\left(\frac{\pi t}{2t_0}\right) , \quad (4)$$

where $\boldsymbol{u}_{1,1}(X_0, Y_0, t)$ is the instantaneous fluid velocity field at the leading order (order $\alpha\beta$),[36] during one scan (i.e., valid for $-t_0 \leq t \leq t_0$), $r \equiv [(x - Ut)^2 + y^2]^{1/2}$ is the distance of the laser spot to the scan path center at time $t$, $A(t)$ is the dimensionless heat-spot amplitude function, $t_0 \equiv l/U$ is half a scan period, and $U$ is the speed at which the laser moves along the scan path. Here, the unit vectors in the $x$- and $y$-directions are denoted by $\boldsymbol{e}_x$ and $\boldsymbol{e}_y$, respectively. The parameters $U$, $t_0$, and $l$ are connected through the scan frequency: $f \equiv 1/(2t_0) \equiv U/(2l)$. The factor of $\frac{3}{2}$ in the net displacement (Equation (1)) reflects the assumption that particles lie in the mid-plane.

The precise position of a particle after a specified duration $\Delta t$ of laser scanning can be calculated by iteratively determining the position following each scan, which is linked to $\Delta t$ via the scan frequency $f$.

In this study, estimates based on previous work were employed for the heat spot radius $a$, the peak temperature change induced by the laser $\Delta T_0$, the thermal expansion coefficient $\alpha$, and the thermal viscosity coefficient $\beta$.[28,36] The scan path length $2l$, the scan frequency $f$, and the scan duration $\Delta t$ were manually predefined in accordance with the specifications of the experiment and simulation. All parameters were maintained at a constant value throughout the duration of a single experimental or simulation run.

Typical values for all these parameters are included in the following table:

| Parameter | Value |
|---|---|
| **Scan path length $2l$** | 15.54 µm |
| **Scan frequency $f$** | 2000 Hz |
| **Scan duration $\Delta t$** | 1.67 s |
| **Heat spot radius $a$** | 4 µm |
| **Temperature change $\Delta T_0$** | 8 K |
| **Thermal expansion coefficient $\alpha$** | 0.0005 K$^{-1}$ |
| **Thermal viscosity coefficient $\beta$** | 0.04 K$^{-1}$ |

In the initial experiments, a discrepancy was observed between the absolute flow speed in experiments and the predicted value from the model using the parameters above. Therefore, the modeled velocity was increased through multiplication with an empirical factor, thereby effectively accelerating the modeled flow field. This yielded a satisfactory agreement between the observed particle displacements in the experiment and the model prediction over the given scan duration, which enabled the successful transfer of the feedback method from simulation to experiments. This empirical factor ranged from 2.4 or 4 in experiments to 9 in simulations. It was speculated that these discrepancies in absolute speed between the model and the experiment can mainly be attributed to 1) deviations in the heat-spot amplitude function and 2) the relatively large particle size compared to the chamber height. The temperature amplitude model (Equation (4)) was based on previous experimental work, capturing modulation of the heating during a laser scan period.[31,35] It was found in the Supporting Information that, instead using an essentially constant amplitude, which corresponds to the relevant experimental conditions for the present work, significantly reduced the empirical factor (Figure S4, Supporting Information). The other major factor contributing to the discrepancies between the model and the experiment is likely the difference in speed experienced by a particle that is large relative to the chamber height (3 µm diameter vs 5 µm chamber height) compared to the point-like particles in the model. From a hydrodynamic perspective, the flow field along the direction perpendicular to the plates has a parabolic profile, with zero velocity on both plates (no-slip boundary condition) and maximum speed at the mid-plane. Therefore, it may be intuitively expected that larger beads would move slower than tracer particles precisely at the mid-plane, as their size means that they experience not only the fastest flow velocities at the mid-plane, but also the lower speeds closer to the no-slip walls. In addition, discrepancies between the assumed values of $\Delta T_0$, $\alpha$ and $\beta$ and the experimental conditions could contribute to this difference in absolute speed.

To accelerate the processing of each individual iteration, an Euler approximation of the precise model was used. The simulated time window, defined by the scan duration $\Delta t$, was divided into multiple linear segments (typically three per scan in experiments), thereby approximating the curved trajectories (Figure S1a, Supporting Information). This resulted in a reduction of the number of calculations required for the forward simulation by approximately three orders of magnitude, consequently reducing the execution time per simulated laser scan path by a similar factor.

**Objective Function Including Interparticle Repulsion**

The objective function $\sigma$ was formulated as a weighted sum of two components: the root-mean-square distance $L$ between particles and their assigned target positions, and the repulsive effective LJ potential $V_{LJ}$ between particles. The potential was employed to prevent particles from colliding in simulations. It does not account for the actual physical interactions between particles. In summary, the objective function is defined by the following set of equations

$$\sigma = \gamma \cdot L + \delta \cdot V_{LJ}, \quad (5)$$

$$L = \sqrt{\frac{1}{n}\sum_{i=1}^{n}|\boldsymbol{p}_i - \boldsymbol{t}_i|^2}, \quad (6)$$

and

$$V_{LJ} = \sum_{i=1}^{n}\sum_{j=i+1}^{n}\begin{cases} 4\varepsilon\left[\left(\frac{2d}{|\boldsymbol{p}_i-\boldsymbol{p}_j|}\right)^{12} - \left(\frac{2d}{|\boldsymbol{p}_i-\boldsymbol{p}_j|}\right)^{6}\right], if\ |\boldsymbol{p}_i - \boldsymbol{p}_j| < 2d \\ 0, \text{otherwise} \end{cases}, \quad (7)$$

with $\gamma$ and $\delta$ denoting the weights, $2d$ the critical interparticle distance, $\varepsilon$ the potential well depth, and $\boldsymbol{p}_i$ and $\boldsymbol{t}_i$ the particle and target coordinates, respectively. In the simulations, the following parameters were used for the objective function including interparticle repulsion:

| Parameter | Value |
| --- | --- |
| **Critical interparticle distance $2d$** | 8.8 µm |
| **Weight of RMS distance $\gamma$** | 0.023 µm$^{-1}$ |
| **Weight of LJ potential $\delta$** | 1 |
| **Potential well depth $\varepsilon$** | 1 |

For simulations without repulsive (effective) Lennard-Jones potential, $\delta$ was set to 0.

**Simulations of Automatic Alignment Using Random Search Algorithm**

In the simulations of the optimization algorithm, the initial particle positions were randomly generated within a 133.2 µm square window according to a uniform distribution. The target locations and critical interparticle distance were predefined manually. The center of the scan path was sampled from a slightly larger square window of 177.6 µm width. The movement of particles due to a flow field selected by the optimization algorithm was simulated in a fully deterministic manner using the analytical flow field model, excluding stochastic effects such as diffusion. A simulation run was terminated when the particles reached the target locations, indicated by the objective function value falling below a predefined threshold, or when the predefined limit on the number of iterations was reached.

In the simulations, five Euler intervals and a scan duration of 3.35 s were employed for each scan path. To accelerate the computational process, a multithreaded approach was employed, whereby multiple processing threads were utilized to evaluate distinct random scan paths in parallel. The simulations were conducted on an 8-core ARM processor with 8 threads, resulting in a calculation speedup of approximately a factor of six compared to a single-threaded implementation.

**Implementation of Random Search Algorithm for Alignment Experiments**

The implementation of the control algorithm in this experimental setup was based on a modification of the hybrid framework for feedback control of thermoviscous flows, which we already described in our previous work.[28,30] In this framework, the image acquisition, particle detection, timing of the feedback loop, and IR laser control were implemented in LabVIEW. The random search algorithm was implemented in Python and interfaced with LabVIEW via the Transmission Control Protocol (TCP). Particles were detected in the camera image using a thresholding segmentation and subsequent calculation of the center of mass.

To reduce the calculation time and achieve a feedback sampling period of 1.67 s, the core flow field evaluation algorithm was executed on ten processing threads simultaneously. The time required to identify a suitable flow field exhibited a wide range, from less than a second to up to 12 s (see Figure S2b in the Supporting Information). This can be attributed to the stochastic nature of the random search algorithm as well as the relatively long time required to model the particle motion due to the flow field (Figure S1, Supporting Information). The long calculation times, in conjunction with the TCP communication framework, necessitated the incorporation of a waiting period between each execution of the control algorithm. During this interval, the system would await the termination of the flow field calculation and refrain from conducting any laser scanning. If the calculation of a flow field exceeded the duration of the feedback sampling period, the waiting period was extended. This was implemented to ensure that flow field calculations were always executed based on the most recent particle positions, and that longer calculations would not be terminated. The image sampling period was controlled independently and was set to 100 ms for the continuous observation of microparticle behavior.

In experiments, particles were located within the camera field of view (88.8 μm squared), while the scan paths were sampled from a larger square window of 177.6 μm width and height, analogous to simulations. Three Euler steps were chosen to approximate the 2D flow field model, thereby enabling a lower feedback sampling period while maintaining reasonable accuracy (see Figure S1a,b in the Supporting Information). The scan duration was aligned with the feedback sampling period, and the length of the scan path was defined manually for each experiment.

For the experiments depicted in Figures 3 and 4, scan path lengths of 17.8 and 14.4 μm, respectively, were utilized, along with required improvements of 10% and 3%, respectively. However, the theoretical displacements for each potential scan path were calculated using a 15.5 μm long scan path. The laser power in the sample plane was determined to be 47.4 mW. The modeled flow speed was increased by empirical factors of 4 and 2.64, respectively. In the experimental setting, the RMS distance was utilized as the sole objective function, without the inclusion of the effective Lennard-Jones potential ($\delta = 0$).

**Theoretical Power-Law Scaling for Characteristic Laser Scan Path Distance to Nearest Particle**

Based on the theoretical model of Liao et al.[36] introduced above, a scaling argument could be used to find a power-law relation between the characteristic distance $r$ of the laser scan path from the nearest particle, and the RMS distance to targets, illustrated in Figure 3e. This characteristic distance $r$ may be interpreted as an approximate upper limit for the laser scan path distance.

The average velocity of tracers induced at the nearest particle has magnitude approximately given by $B/r^2$, provided the scan path is sufficiently far away, where $B$ is a prefactor given by the theory as[36]

$$B = \frac{3}{2} \times \frac{1}{8t_0} \alpha\beta\Delta T_0^2 U a^2 \int_{-t_0}^{t_0} A(t)^2 \, dt. \quad (8)$$

After the scan duration (time interval) $\Delta t$, the nearest particle has characteristic net displacement $B\Delta t/r^2$.

In one step, the RMS distance to targets, denoted here by $L$, must reduce by $pL$, where $p$ is the required improvement as a proportion, satisfying $0 < p \leq 1$. The characteristic distance r of the laser scan path from the nearest particle may therefore be found as

$$r = (B\Delta t/p)^{1/2} L^{-1/2}. \quad (9)$$

This characteristic $r$ gives an approximate upper bound for the distance of the laser scan path from the nearest particle, as particles further away move by a smaller distance than $pL$.

To explain the experiments in Figure 3, the parameter values in the following table were used:

| Parameter | Value |
|---|---|
| Required improvement $p$ | 10% |
| Scan path length $2l$ | 15.54 µm |
| Scan frequency $f$ | 2000 Hz |
| Scan duration $\Delta t$ | 1.67 s |
| Heat spot radius $a$ | 4 µm |
| Temperature change $\Delta T_0$ | 8 K |
| Thermal expansion coefficient $\alpha$ | 0.0005 K⁻¹ |
| Thermal viscosity coefficient $\beta$ | 0.04 K⁻¹ |

and, further, the prefactor $B$ was replaced with $4B$ to calibrate the maximum speed to experiments (empirical factor, as discussed together with the analytical model details).

**Microparticle Sample Preparation for Experiments**

For the alignment experiments, chambers containing probe particles immersed in a thin liquid film were constructed. This procedure was highly similar to that used in the previous work.[28,30] The medium was a 60% (v/v) glycerol–water mixture (glycerol puriss., 15523-M, Sigma–Aldrich). The probe particles utilized in this study were spherical green, fluorescent PS particles of 3 µm diameter (Fluoresbrite YG Microspheres, mean diameter 3.06 µm, 171552, Polysciences). To prevent the PS particles from adhering to the glass walls, 0.3% v/v Tween 80 (P1754, Sigma–Aldrich) was added to the liquid. The film thickness was reproducibly approximately 5 µm when a defined volume of liquid was used between the slides (2 µL), and additional fluorescing particles of 5 µm diameter were added to the mixture (based on melamine resin, FITC-marked, 75908-5ML-F, Sigma–Aldrich). The minimal thickness of the film prevents the probe particles from exiting the imaging plane during the alignment process. To enable the measurement of chamber height using confocal scans of the chamber along the z-axis, low concentrations of 0.1 µm diameter fluorescing PS particles (fluorescent carboxyl polymer particles, mean diameter 0.11 µm, FC02F, Bangs Laboratories, Inc.) were additionally added to the mixture. To create the chambers, the mixture was sandwiched between a microscope objective slide and a cover glass (Ø 22 mm). To prevent drift, the sides of the chambers were sealed using dental impression material (Identium Light, Kettenbach Dental).

**Use of Scientific Software**

ImageJ Fiji[47] was used to process microscopy images and generate overlays of laser scan and target positions.

**Conflict of Interest**

E.E. and M.K. applied for an international European patent for optofluidic technology related to this publication (application number PCT/EP2021/071437). M.K. is a co-inventor of technology for laser induced flow and force measurement technology (US Patent App. 17/506,750 and PCT/EP2021/071392) and holds a consultancy contract with Rapp Optoelectronic GmbH.

**Author contributions**

IS and MK wrote the random search simulations code. WL and EL developed the mathematical model and code for the thermoviscous fluid flow. IS generated simulation data. EE and IS performed the experiments. EE, IS and MK analyzed the data. MK, EE and IS conceived the project. MK wrote first draft, which EE, IS, WL, EL and MK finalized with input from FN. All authors participated in the critical discussion of the final draft.

**Data Availability Statement**

The data that support the findings of this study are available from the corresponding author upon reasonable request.

**Acknowledgements**

E.E. and I.S. contributed equally to this work. The authors acknowledge funding by the Karlsruhe Institute of Technology and the University of Cambridge. M.K. further acknowledges support by the European Research Council, in particular the ERC Starting Grant GHOSTs (Grant No. 853619) and the Hector Foundation. M.K. and E.E. acknowledge support by the Volkswagen Foundation (Life! Grant No. 92772). Additionally, the Kreysing lab was co-funded by the Deutsche Forschungsgemeinschaft (DFG, German Research Foundation) under Germany's Excellence Strategy—2082/1–390761711 (3DMM2O), Research Grant 515462906, and the DFG project 7593. The authors also acknowledge support by the Karlsruhe

School of Optics & Photonics (KSOP) and the Helmholtz Program Natural, Artificial and Cognitive Information Processing (NACIP). W.L. gratefully acknowledges funding from the Engineering and Physical Sciences Research Council (EPSRC studentship) and Trinity College, Cambridge (Rouse Ball and Eddington Research Funds travel grant). The authors again thank their colleagues for their valuable feedback and discussions. They thank previous members of the research group for establishing the fundamental framework for the experimental setup and control software for thermoviscous flows.

The authors employed ChatGPT4o for the purposes of coding assistance (autocomplete and code structuring, annotation), as well as for tone adjustment, proofreading, and feedback on writing style. Additionally, the authors used DeepL for the purposes of proofreading and language improvement.

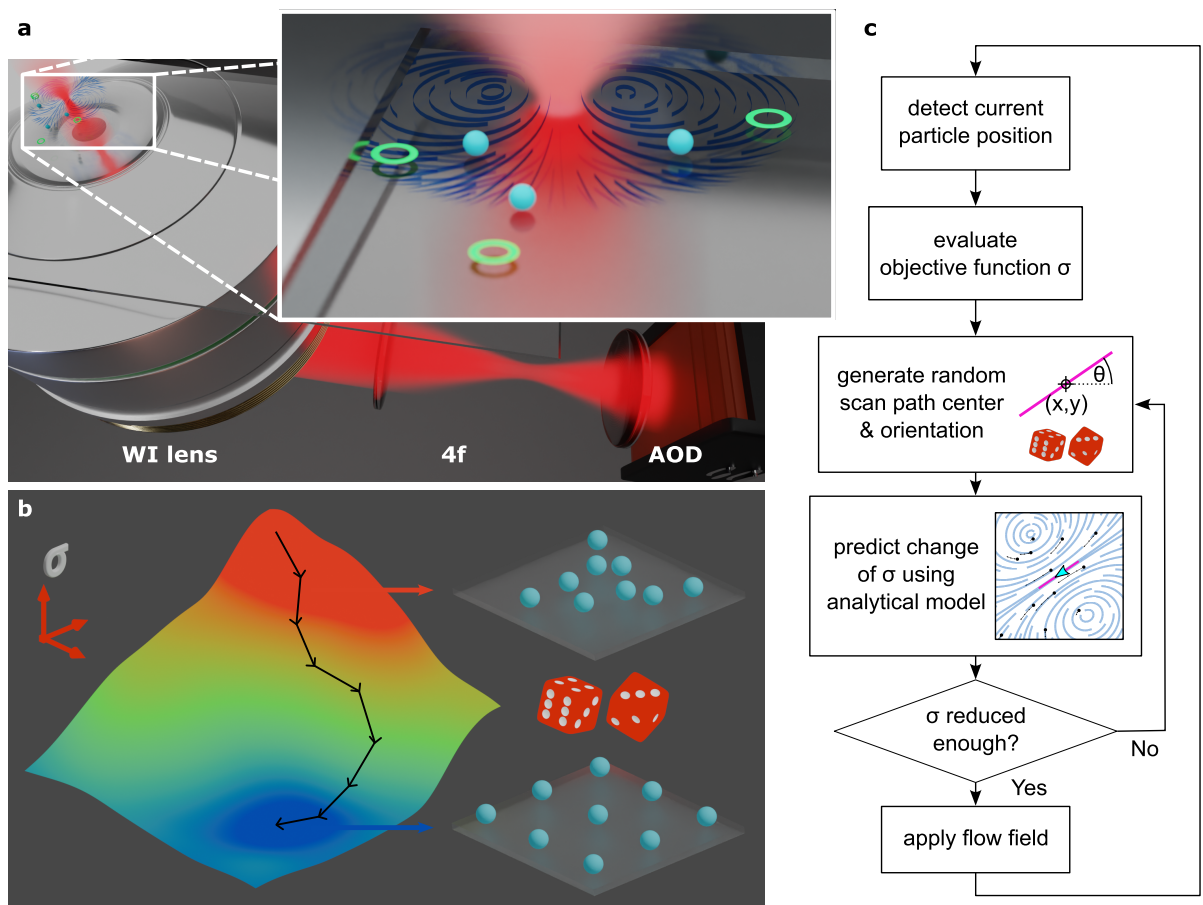

**Figure 1: Thermoviscous microparticle assembly via model predictive control.**
a) The key components of the optofluidic setup, consisting of an infrared laser beam (1455 nm) that is scanned across a sample chamber via a 4f telescope, a two-axis acousto-optic deflector (AOD), and a water immersion (WI) lens. The magnified view shows particles being guided toward their intended target destinations by suitable laser-induced flow fields.
b) The alignment of particles is achieved through stochastic optimization of an explicit objective function ($\sigma$). This is facilitated by stepwise random search.
c) The flowchart depicts the closed loop implemented in the simulation and experiments to select suitable flow fields using random search methodology. The algorithmic process is as follows: 1) evaluate the objective function $\sigma$ based on the current positions of the particles and the targets; 2) randomly generate a new location and orientation for the scan path; 3) use an analytical model to predict the updated $\sigma$ after applying the proposed flow field; and 4) accept the new scan path if the predicted $\sigma$ meets the requirements, otherwise repeat steps 2–3.

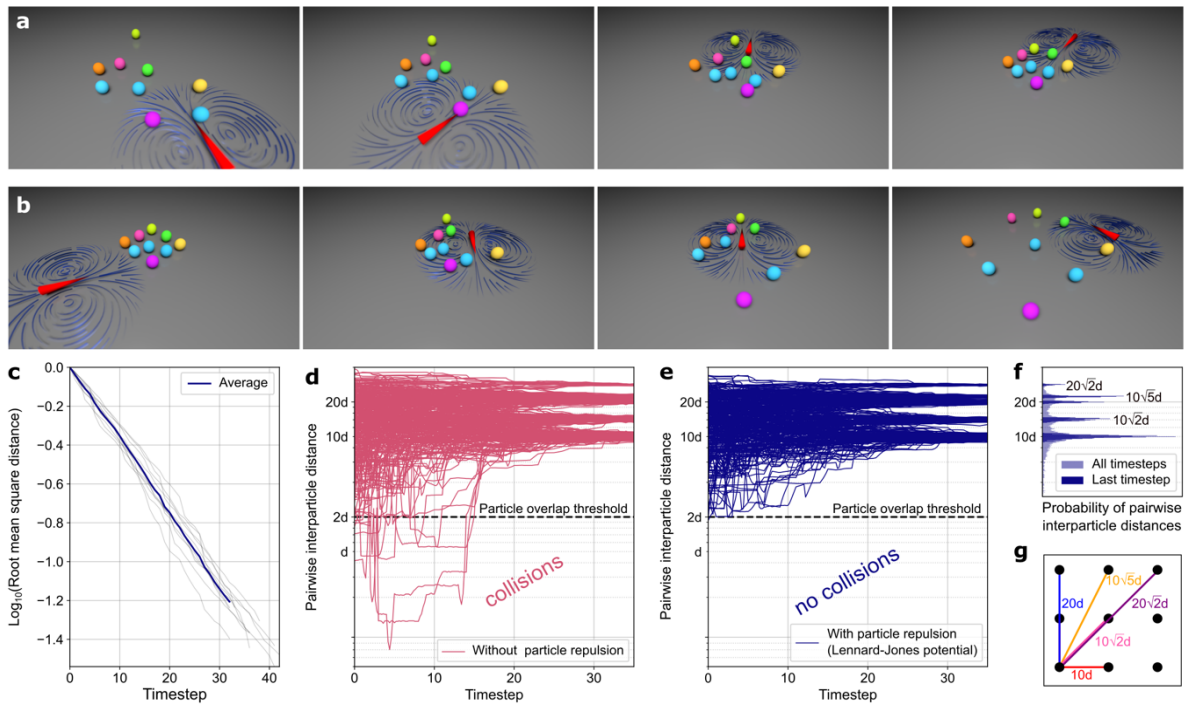

**Figure 2: Deterministic simulations showing sufficiency of thermoviscous flows to align microparticles.**

a) 3D renderings of exemplary data from stepwise deterministic simulations of the approach and alignment of nine particles to form a grid-like target arrangement using stochastic optimization. The root-mean-square (RMS) distance between the particles and their designated targets serves as the objective function. For improved visual differentiation, the particles are displayed in various colors. In the simulations, particles are physically indistinguishable and are not assigned an identity.

b) Expansion of the 3 × 3 grid to an edge-to-edge distance of 10 times the critical interparticle distance ($2d$, see the "Methods"). A video recording of sample data showing the approach and subsequent expansion of the grid is included as Video S1 (Supporting Information).

c) The objective function (root-mean-square distance to targets) decreases monotonically over time in a nearly exponential manner for the scenario in panel (a). Data from ten repetitions are shown in gray, and the average is shown in dark blue.

d) The interparticle distances are shown during alignment simulations of the 3 × 3 grid before the introduction of pairwise repulsive particle interactions. In the absence of particle repulsion, violations of excluded volumes are evident. Data curves from ten repetitions are shown, and the pairwise interparticle distance is plotted logarithmically ($y$-axis).

e) The alignment simulations were repeated with the addition of particle repulsion to the objective function, implemented through a Lennard–Jones potential. The data curves from ten repetitions clearly show that the excluded volumes are preserved, in contrast to panel (d). The pairwise interparticle distance is plotted logarithmically ($y$-axis).

f) Probability distribution of pairwise interparticle distances under the influence of particle repulsion. In light blue, the distances across all time points from panel (e) are displayed. In dark blue, only the data from the last five time steps are shown. This clearly shows that particles settled into distances corresponding to the 3 × 3 grid.

g) Schematic of the discrete relative pairwise distances present in a 3 × 3 grid of particles.

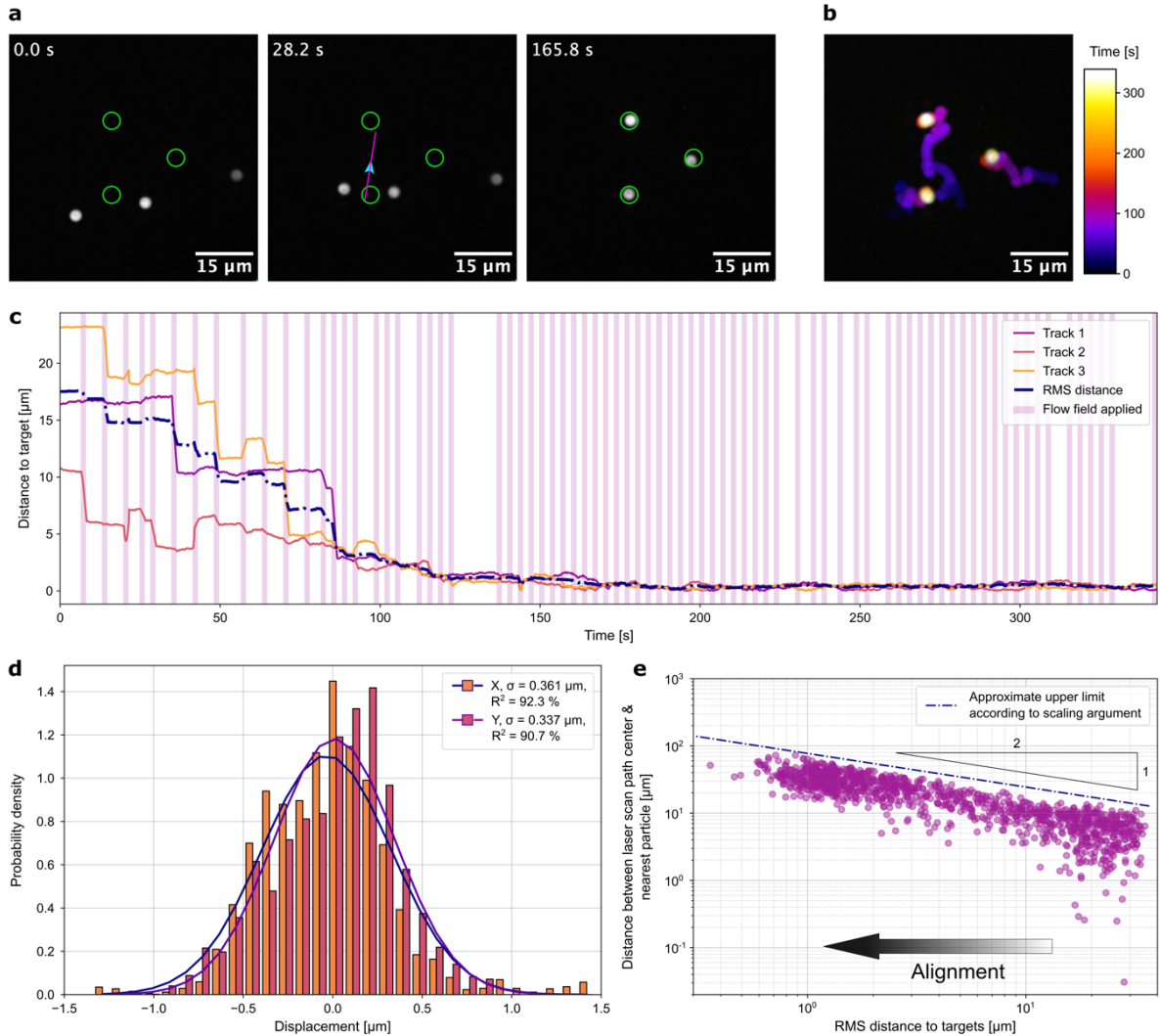

**Figure 3: Proof-of-principle microparticle alignment and emergent remote manipulation.**

a) Representative snapshots of the experimental particle alignment process are shown. The green circles represent target positions; the magenta line indicates the scan path; and the cyan arrow denotes the direction of the center-line flow. A video recording of the complete alignment process is available as Video S2 (Supporting Information).

b) The time-dependent evolution of the particle alignment demonstrates the significant impact of stochastic contributions resulting from diffusive motion.

c) The RMS distance between particles and their target positions, which serves as the objective function, exhibits near-monotonic convergence and excellent stability. It is noteworthy that the distance of an individual particle from its target may temporarily increase.

d) The histogram of the positional distribution of all particles relative to their targets demonstrates a positioning precision of approximately 0.35 µm (rounded) post convergence.

e) The distance between the center of the laser scan path and the nearest particle demonstrates an inverse, nonproportional relationship with the RMS distance. This indicates that upon reaching the RMS distance convergence, the selected scan paths operate at a greater distance from the particles, enabling precision positioning. The presented data points were obtained from ten repetitions of the experiment shown in panels (a)–(d). The detailed time progression of the objective function and the distance between the scan path and the nearest particle for these ten repetitions can be found in Figure S3 (Supporting Information). A theoretical scaling argument yields an approximate upper bound for the distance $r$ from the scan path to the nearest particle, as a function of RMS distance $L$, with power-law relationship given by $r \propto L^{-1/2}$ (full formula in the "Methods"). The laser scan path must be close enough to generate sufficient net displacement of particles that produces the required improvement in RMS distance.

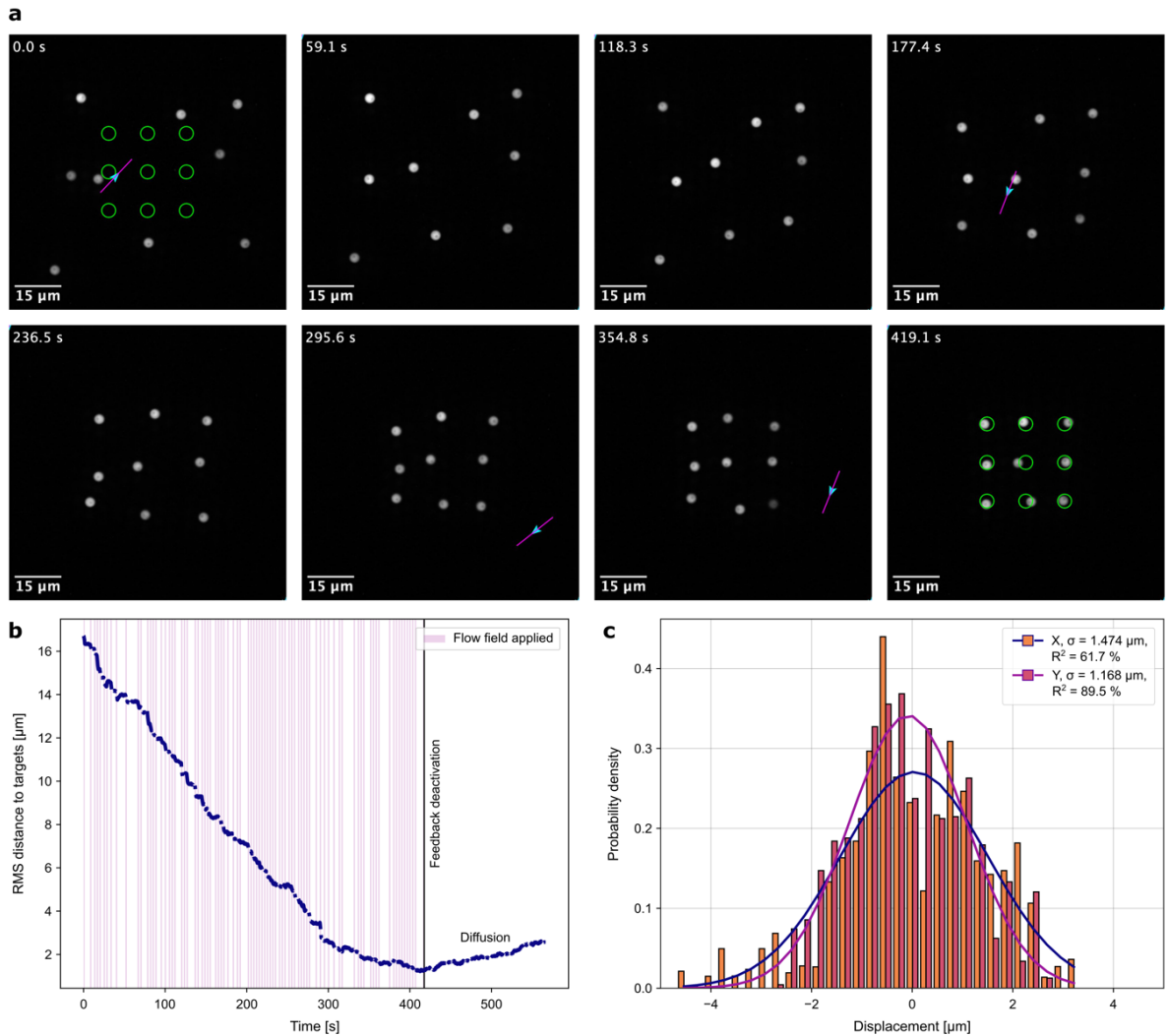

**Figure 4: Simultaneous particle alignment in a crowded setting leveraging stochastic optimization.**

a) Representative sequential snapshots illustrating the alignment of nine colloidal particles to form a 3 × 3 grid analogous to simulations. A video recording of the complete alignment process is available in Video S3 (Supporting Information).

b) The time-dependent evolution of the objective function (root-mean-square distance) exhibits near-monotonic convergence until the end of the experiment. A noticeable increase in RMS distance due to diffusive motion is observed following the deactivation of the flow field application.

c) A histogram displaying the positional distribution of all particles relative to their targets reveals a positioning precision ranging from 1.2 to 1.5 µm (rounded) post convergence.